\documentclass[fleqn,10pt]{wlscirep}
\usepackage[utf8]{inputenc}
\usepackage[T1]{fontenc}

\title{In-situ multicore fibre-based pH mapping through obstacles in integrated microfluidic devices}

\author[1,*]{Harikumar K. Chandrasekharan}
\author[1,2]{Krystian L. Wlodarczyk}
\author[1]{William N. MacPherson}
\author[2]{M. Mercedes Maroto-Valer}
\affil[1]{Applied Optics and Photonics Group, School of Engineering and Physical Sciences, Heriot-Watt University, Edinburgh EH14 4AS, UK}
\affil[2]{Research Centre for Carbon Solutions, School of Engineering and Physical Sciences, Heriot-Watt University, Edinburgh EH14 4AS, UK}

\affil[*]{hk47@hw.ac.uk}

\begin{abstract}
Microfluidic systems with integrated sensors are ideal platforms to study and emulate processes such as complex multiphase flow and reactive transport in porous media, numerical modeling of bulk systems in medicine, and in engineering. Existing commercial optical fibre sensing systems used in integrated microfluidic devices are based on single-core fibres, limiting the spatial resolution in parameter measurements in such application scenarios. Here, we propose a multicore fibre-based pH system for in-situ pH mapping with tens of micrometer spatial resolution in microfluidic devices. The demonstration uses custom laser-manufactured glass microfluidic devices (called further micromodels) consisting of two round ports. The micromodels comprise two lintels for the injection of various pH buffers and an outlet. The two-port system facilitates the injection of various pH solutions using independent pressure pumps. The multicore fibre imaging system provides spatial information about the pH environment from the intensity distribution of fluorescence emission from the sensor attached to the fibre end facet, making use of the cores in the fibre as independent measurement channels. As proof-of-concept, we performed pH measurements in micromodels through obstacles (glass and rock beads), showing that the particle features can be clearly distinguishable from the intensity distribution from the fibre sensor.

\end{abstract}

\begin{document}

\flushbottom
\maketitle
\thispagestyle{empty}

\section*{Introduction}

Microfluidics has emerged as an ideal tool for the precise manipulation and study of small amounts of fluids (nanolitres and picolitres) through channels with dimensions on the scale of tens to hundreds of micrometers. Since it's first application in analysis \cite{MANZ1992253} microfluidics has gained enormous significance and currently find applications in drug delivery,  biology, tissue engineering, diagnostics, cell analysis, virology, microreactors, lab-on-a-chip, and in microrobotics \cite {RIAHI2015101,doi:10.1146/annurev.bioeng.4.112601.125916,BRATTLEAL20137227,bioengineering6040109,Avesar20142161,https://doi.org/10.1002/rmv.2154,C7LC00800G,Filippi}. The key advantage of microfluidic systems compared to traditional macroscale laboratory techniques is that the relatively small dimensions of microfluidic chips allows the use of small volume of samples for reagents mixing, which significantly fasten the reaction time, reduces the reagent cost, and enables real-time monitoring of various physical and chemical processes. In most of these applications, in-situ measurements of parameters such as pH, chemical concentration, temperature, pressure, cellular force, and conductivity are essential. For example, in lung cancer research, microfluidic based fluorescence measurement is used to monitor the growth of human lung adenocarcinoma cell lines, by measuring the cell adhesion forces \cite{Li}. Simultaneous and accurate quantitative measurements of pH, O\textsubscript{2}, and CO\textsubscript{2} in human saliva are possible with low-cost integrated microfluidics devices \cite{1390568838433487616}. In organ-on-chip devices, continuous and real-time monitoring of pH and O\textsubscript{2} levels in a cell culture medium is demonstrated with a cost-effective microfluidic bioreactor \cite{10.1063/1.4955155}. 

Microfluidic devices are often used as effective physical micromodels of porous media to study fluid flow and behaviour at the pore level (microscale level)\cite{Gogoi,s20144030}. For instance, in CO\textsubscript{2} storage and petroleum engineering, microfluidic devices are extensively used to investigate the displacement of fluids, precipitation, reactions of solvents and chemical compounds\cite{C7LC00301C,s20144030,2017WR020850,2019WR025420,es204096w}. The porous structures (patterns) in such microfluidic devices can be created using photolithography, etching, direct laser writing, and various 3D printing techniques, including extrusion-based printing, material jetting, and stereolithography\cite{s20144030,vzj2011,mi12030319,Wlodarczyk,ZHAO2023116864,SuW,D2LC01177H,mi9080409}. The dimensions of porous features in such devices are on the scale of tens of micrometers, offering fine control on the functionality of the devices\cite{SuW,mi9080409, D1LC01086G, Zzzz,pnas.1603387113,PRADHAN2019704}. The functionality and the value of the devices can be further enhanced by integrating the devices with custom-made and/or commercially available sensors\cite{1390568838433487616,s21227493}. When sensors (O\textsubscript{2}, CO\textsubscript{2}, pH, pressure, temperature) are integrated into patterned microfluidic devices, it is important to match the spatial resolution of the sensors with the pattern dimensions. This is particularly beneficial for applications in which the location (spatial coordinates) mapping of the parameter is important (e.g. local pH mapping during protein transport in bacteria\cite{mbio.01911-16}, pH distribution in microfluidic channels \cite{Kyosuke}, or microbial cultivation in droplets\cite{Tovar}). Existing commercial systems, however, are mostly based on single-point measurements which cannot offer satisfactory spatial mapping capabilities within the sensor's active region. For instance, commercially-available fibre-based sensors can be integrated with microfluidic devices to measure O\textsubscript{2}, CO\textsubscript{2}, pH, and pressure\cite{1390568838433487616,s21227493,app11052049, 10.3389}. An optical signal interrogation system is used to collect the signal from the sensors during measurements, which is then interpreted by the sensor's readout unit to provide meaningful data. Typically, in spot sensors (O\textsubscript{2}, CO\textsubscript{2}, pH), the signal collection (and field illumination) from the sensor is achieved using a large single-core fibre within the interrogation unit, which collects the emitted signal across the sensor region. The spatial information in such a system is compromised of a readout unit that interprets the data based on the total emitted signal intensity.

In this work, we developed a pH sensing system using a polymer multicore fibre (MCF), that was integrated with commercial sensors within microfluidic devices. Compared to commercial optical fibre interrogation systems that are based on single-core fibres, our pH sensing system allows pH measurements from 13,000 independent fibre channels. We showcase that when MCF is integrated with a commercial sensor spot (PyroScience GmbH), our system is capable of spatially mapping the pH from the sensor's active area. The sensors are embedded within the microfluidic device using custom-made 3D connectors manufactured using a stereolithography 3D printer\cite{s21227493}. The connectors are designed in such a way that the sensors can be safely connected and removed without damaging the microfluidic device under test. The developed MCF system was compared with a commercial single-core fibre interrogation system for the real-time monitoring of process conditions (pH changes during the mixing of various solutions), which can also be used to predict the MCF system's performance. We propose that the developed system can be used in measurements in porous media micromodels with embedded obstacles for the pH mapping of fluids. To demonstrate this, we emulate a porous medium by embedding microbeads and micro-rocks (rock particles with tens of micrometre dimensions) within the selective ports of a microfluidic device. Exploiting the imaging capability of the polymer MCF, we recorded the emission profile from the sensor spot with a clear visual of the obstacles that are in close contact with the sensor. In commercial single-core fibre pH systems, the presence of obstacles would result in the reduction of the emitted signal, which would lead to a measurement error as the readout unit generates the pH value based on total signal intensity. 

\section*{Results and discussion}
\subsection*{Experimental system}

\begin{figure}[b!]
\centering
\includegraphics[width=\linewidth]{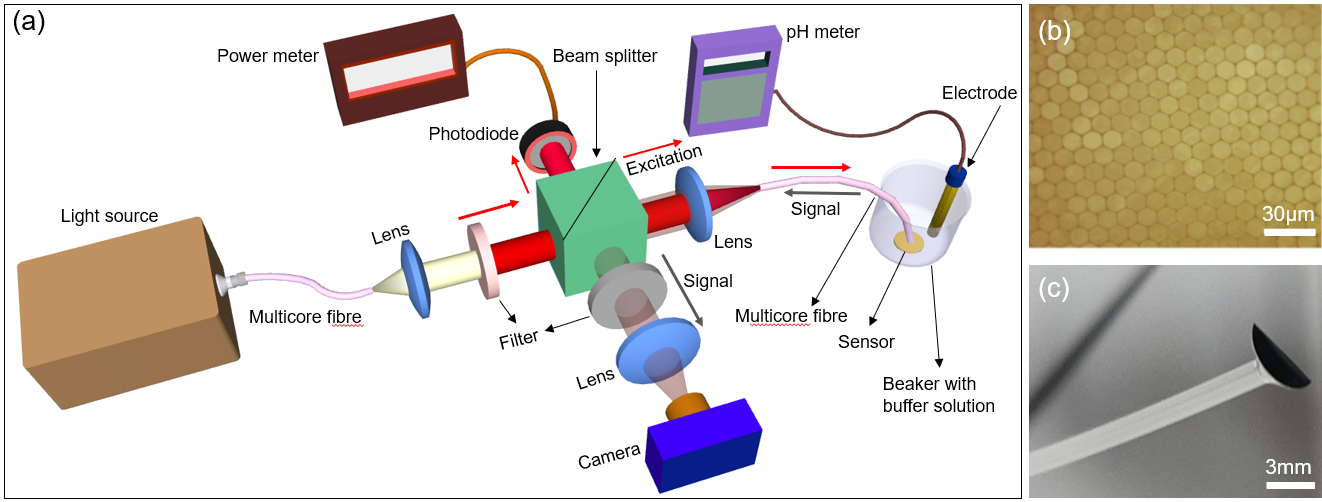}
\caption{(a) Experimental system used for the MCF-based pH sensing. A beamsplitter was used to send the excitation light to the MCF and the reflected fluorescence signal is sent to the imaging system using the reflection arm of the beamsplitter. \newline(b) A micrograph of the MCF facet. (c) Photograph of the MCF with the sensor spot attached.}
\label{fig:1}
\end{figure}

Figure \ref{fig:1}a shows the schematic of the experimental system used for the multicore fibre-based pH mapping. A fibre illuminator (Thorlab OSL-1-EC) provided a broadband white light source, which was coupled to a short length of a polymer MCF (MBL-1500 (13000 cores, diameter 1.5mm),  Asahi Kasei Corporation). The emitted light from the fibre is delivered to an aspheric lens (f= 25 mm) which collimates the beam. A narrow passband filter was used to select the appropriate excitation wavelength (Central Wavelength- 626 nm, Full-width Half Maximum- $\pm$ 13 nm, Ealing) for a pH sensor spot (Pyroscience GmbH). The selected excitation light was then focused onto the input of a 20 cm long polymer multicore fibre (MBL-1500,  Asahi Kasei Corporation) using a second aspheric lens (f = 50 mm). The 
reflected light from the beam splitter is sent to a power meter (Thorlabs PM100D) to monitor the input light intensity. The variation in source light intensity was found to be $\pm$ 2.5\% to the mean value. The polymer MCF exhibited a diameter of 1.5 mm and consisted of 13,000 individual cores (Figure \ref{fig:1}b). Each core has a 12 $\mu$m diameter and a numerical aperture of 0.5. Both facets of the MCF were hand polished normal to the fibre axis to avoid losses from the facets due to imperfect cleaves. The sensor spot contains a pH-sensitive coating on a transparent substrate which is attached to the distal end facet of the MCF with transparent silicon adhesive (room-temperature curing adhesive recommended by the manufacturer) as shown in Figure \ref{fig:1}c. In this geometry, the sensor coating is in direct contact with the medium under test.  The silicon glue does not affect the light transmission into and from the sensor spot and is left to set for 12 hours in the dark before the measurements. The fluorescence-based sensor spot (PHSP5-PK7, PyroScience GmbH) was designed to work in the pH range 6-8, but can be used beyond the specified range for calibration purposes\cite{link}. The sensor spots are designed to change the fluorescence emission intensity when submersed in different pH buffers. Very briefly, the sensor consists of a pH-insensitive reference indicator and a pH-sensitive luminescent
dye. Both are excited with red light (at a wavelength of 610 - 630 nm) and show bright luminescence in the near-infrared (760-790 nm). The MCF (with the sensor spot attached) was then secured on a linear translation stage (not shown) which allows the fibre to be submersed and retracted into different pH solutions. The emitted backscattered fluorescence light from the sensor is coupled onto the optical fibre and is then incident on the input facet of the fibre sensor. The reflected light is collected by the reflection arm of the beamsplitter and sent to a  confocal imaging system. The confocal system consists of an aspheric lens (f= 35 mm) which images the input facet of the MCF, an emission filter (Central Wavelength- 780 nm, Full-width Half Maximum- $\pm$ 10 nm, Ealing), and a CMOS camera (Thorlabs DCC1545M-GL). The emission filter was used to block the stray light and the reflected excitation light from the input facet of the MCF. The coupling of light between the cores of the MCF is negligible for such a short length of MCF at 780 nm\cite{Chandrasekharan}. This means the cores of the MCF can be used as independent measurement channels and the MCF will preserve the spatial profile of the back-reflected fluorescence emission from the sensor.

\subsection*{Calibration}

 The first step in the calibration of the imaging fibre system was to estimate the response time of the sensor for a rapid change in the pH environment. For this, the sensor spot was dipped into two different standard pH buffers (purchased from Thistle Scientific Ltd) at room temperature. First, the sensor spot was submersed in a buffer solution of pH 8.13 and the intensity was recorded in a real-time manner on the CMOS camera. The fibre sensor was then retracted from the solution and re-immersed in a buffer solution of pH 7.46. The change in emission intensity was monitored on the camera at a frame rate of 1 frame-per-second (see supplementary Visualisation 1). The time taken to stabilise the fluorescence emission upon the rapid change in pH was calculated from the total intensity of the image frames. A signal rise time of 3.4 $\pm$ 2 seconds was calculated from the total counts of the intensity image frames recorded on the camera (normalised to the total counts in the image frame with the maximum counts, see Supplementary Figure S1a). The rise time is defined as the time taken for a 90 \% increase from the initial pH value to the final pH value. The measurement is repeated to further evaluate the response time in the recommended calibration range of the sensor. This time, the measurements were conducted with buffers of pH values of 2 and 11. Initially, the sensor was submersed in a pH 2 buffer for 1 hour to stabilise, and the emission profile of the sensor was recorded on the camera. The sensor was then retracted from the pH 2 buffer and immediately submersed in the pH 11 buffer. Four images were then taken at 30 seconds, 5 minutes, 13 minutes, and 25 minutes, to evaluate the response time of the sensor in reverse direction. The total counts for the five images (one image for pH 2 and four images for pH 11) were plotted as the function of time (see supplementary Figure S1b) and the signal fall time was evaluated from the exponential fit to the data. The fall time of the sensor (defined as the time taken for a 90 \% decrease from pH 2 value to pH 11) is calculated to be 12 s. Before the intensity measurements, the pH values of the buffers were measured using a calibrated commercial pH meter (HI-2002 Edge pH meter, Hanna Instruments UK) with a measurement accuracy of $\pm$ 0.01.
  \begin{figure}[t!]
\centering
\includegraphics[width=0.7\linewidth]{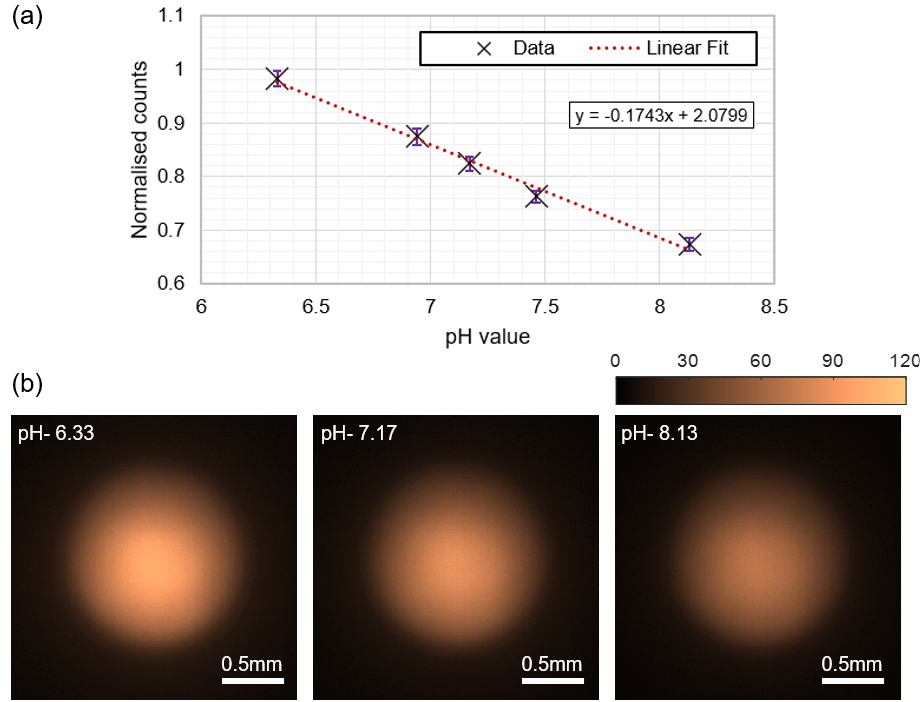}
\caption{(a) Calibration curve for the MCF sensor and the linear fit of data for 5 different pH solutions measured at room temperature with the error bars. (b) False color images of the recorded spatial fluorescence intensity distribution from the MCF sensor for three different pH buffers.}
\label{fig:2}
\end{figure}

 Once the response time was estimated, calibration measurements were performed for the sensor. Five pH buffers were prepared using commercially available buffer solutions with pH values 6.33, 6.94, 7.17, 7.46, and 8.13 (sensor spot pH range 6-8), measured using the Hanna Edge pH meter. The pH measurements were repeated 5 times, and the standard deviation was found to be $\pm$ 0.01 (which matches the accuracy of the pH meter). The pH values of the selected buffers from the manufacturer data sheet were 6.4, 6.8, 7, 7.4, and 8 respectively. The ionic strength of the pH solutions was set to 150 mM, as recommended by the sensor's manufacturer. The pH values of the buffers were measured with the commercial pH meter. As shown in Figure \ref{fig:1}, the sensor was submersed in the prepared solutions and the fluorescence emission profile was recorded on the CMOS camera for the five buffer solutions. For each pH intensity image, the sensor was submersed in the solution for five minutes to stabilise the sensor's fluorescence emission (Note that, this is at least 25 times higher than the response time of the sensor, measured previously). The fluorescence emission profile was recorded on the CMOS camera for each pH buffer ten times, and the total signal counts were plotted as the function of pH value. The fluorescence signal (total counts in the image) varies across the fibre cores for different buffer solutions, which directly indicates the change in the pH environment. Figure \ref{fig:2}a represents the calibration curve of the sensor spot for pH values 6.33, 6.94, 7.17, 7.46, and 8.13 and the linear fit to the data. Based upon the noise observed in the intensity measurements, the pH resolution of the MCF sensor was found to be $\pm$ 0.16 (95 \% confident). The total intensity values among 10 sets of experiments vary $\pm$ 2.5\% which indicates that the uncertainty in the intensity measurements is dominated by the input light source power variations. Figure \ref{fig:2}b represents three intensity images of the fibre sensor for buffers with pH values of 6.33, 7.17, and 8.13 respectively. Note that, due to input light power limitation, we were not equally exciting the sensor spot, due to the intensity gradient observed in Figure \ref{fig:2}b.

\subsection*{Spatial pH mapping using a multicore fibre sensor}
\begin{figure}[b!]
\centering
\includegraphics[width=0.6\linewidth]{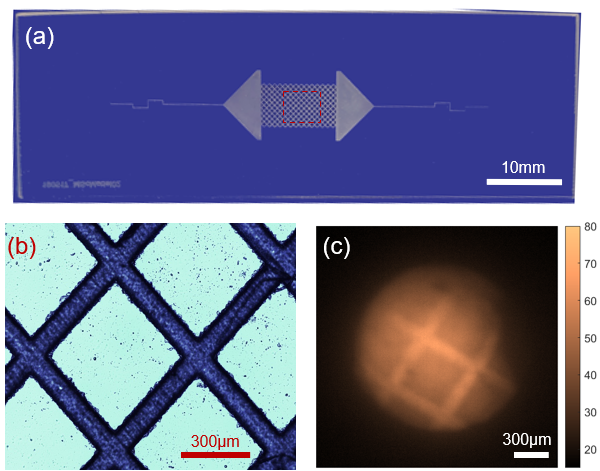}
\caption{(a) A photograph of the microstructured device used for pH spatial mapping analysis of the sensor. The rectangular region at the centre represents the approximate area in which the fibre sensor was kept in close contact with the micromodel. \newline(b) Micrograph of the device with laser-fabricated channels. (c) The recorded intensity image on the camera shows well- distinguished microchannels.}
\label{fig:3}
\end{figure}

A laser-generated microfluidic pattern on a glass substrate is used to evaluate the fiber sensor's spatial pH imaging capability. This pattern is chosen due to it's well defined structure on a scale appropriate for a typical pore-scale micromodel. As shown in Figure \ref{fig:3}(a)\&(b), the glass sample consisting of square array channels of width of 100 $\mu$m and depth of 100~$\mu$m, fabricated using ultrafast laser micromachining \cite{Wlodarczyk,mi9080409}. The micro-patterned sample (or micromodel) is manufactured from a borosilicate glass plate (Borofloat33, SCHOTT Technical Glass Solutions GmbH, Germany) with dimensions of  75~mm~×~25~mm~×~1.1~mm. The micromodel is submersed in a Petri dish filled with a buffer solution of pH 8.13. The sensor spot was then translated and kept in close contact with the micro-structures in the model using the translation stage. When the sensor spot is in close contact with the micromodel, we expect higher signal emission from the sensor regions which are in close contact with the channels. This is due to the fact that the sensor spot is exposed to more fluid in the microchannel compared to the regions in the micromodel outside the microchannels. The emission profile from the sensor is taken after 5 minutes of sensor exposure on the CMOS camera. As the coupling of light between the MCF cores are negligible, the emission profile from the sensor spot is preserved when travels back to the input facet of the MCF. The channel structures can be clearly visible in the fibre intensity images as shown in Figure \ref{fig:3}c. It can be noted that regions outside the microchannel emit light as the sensor submersed in the fluid before coming into close contact with the micromodel.    

\subsection*{Comparison of MCF sensor with single-core commercial fibre interrogation system}
\begin{figure}[http]
\centering
\includegraphics[width=0.7\linewidth]{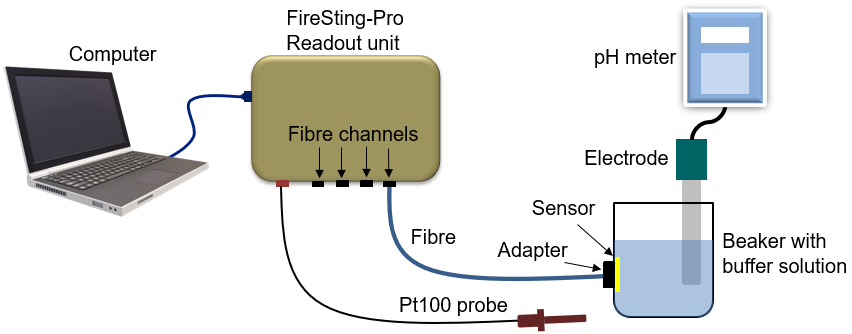}
\caption{Experimental setup used for FireSting-Pro fibre interrogation system calibration and sample measurements.}
\label{fig:4}
\end{figure}

To evaluate the pH sensing capabilities of the developed pH imaging probe, we demonstrated the MCF pH system alongside  a commercial single-core fibre interrogation system. The commercial single-core fibre system (FireSting-Pro, Pyroscience GmbH) comprises four fibre channels, with independent pH sensing capabilities with a specified accuracy of $\pm$ 0.01 and a temperature probe (Pt100 probe) for automated temperature compensation during the measurements\cite{link1}. The fibres used here are single-core polymer fibres with a core diameter of 1 mm that can be used in combination with the sensor spots. The fibres are used to transmit the excitation light (Central Wavelength- 625 nm, Full-width Half Maximum- $\pm$ 7 nm) to the sensor spot and to collect the backscattered fluorescence emission (Central Wavelength- 780 nm, Full-width Half Maximum- $\pm$ 10 nm) from the sensor spots. The interrogation readout unit then provides the pH value of the solution based on the fluorescence emission intensity. Before the measurements, a two-point calibration is performed at room temperature with recommended pH calibration buffers of nominal pH values 2 and 11 from the manufacturer. As shown in Figure \ref{fig:4}, the sensor spot is attached to the inner sidewall of a 50 mL beaker with silicone gel, and the beaker is filled with pH calibration buffers. In order to hold the fibre close to the sensor, a fibre adapter is attached to the outer side wall using an adhesive. The fibre is then inserted into the adapter and secured using a set screw, enabling illumination and readout from the sensor. For each measurement, the FireSting-Pro readout unit provided real-time pH values, and the acquisition is continued until a comparison pH value (flat line) was obtained from the readout software. 

\begin{figure}[t!]
\centering
\includegraphics[width=\linewidth]{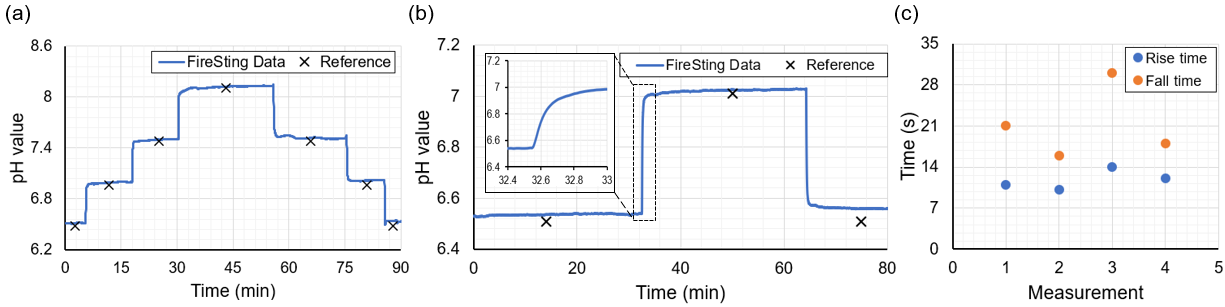}
\caption{pH response behavior of the sensor (a) pH values have been staircase changed from 6.48 to 8.13 and then back to 6.48. (b) The response curve of the sensor for pH 6.48 to 6.96 and then back to pH 6.48. The black symbol indicates the measured pH value using the reference pH meter. (c) The rise time and fall time of the sensor for 4 cycles of measurements.}
\label{fig:5}
\end{figure}

 Following calibration, pH buffer solutions were prepared with ionic strength set at 150 mM from stock pH buffers. Four pH buffer solutions were prepared with pH values 6.48, 6.96, 7.48, and 8.13, measured using a calibrated reference pH meter (Hanna Edge pH meter) with accuracy $\pm$ 0.01. As shown in Figure \ref{fig:4}, during pH measurements, the pH values of the buffer solutions were accurately monitored with the reference pH meter with the same pH measurement accuracy. Both systems (PyroScience and Hanna Edge pH meter) showed similar pH values for four pH buffers with a standard deviation of $\pm$ 0.04, measured from a set of 5 measurements. The response behavior and the real-time pH sensing capability of the sensor upon a rapid change in pH solutions were evaluated. For this, the pH has been staircase changed from 6.48 to 8.13, and vice versa, and the sensor response is monitored using the real-time FireSting readout unit. Figure \ref{fig:5}a, presents the sensor behavior for the forward and backward staircase pH change. The pH buffers from the beaker were quickly changed from the beaker (within 10 seconds) manually during each buffer change. During this change of fluids, the sensor was still wet and the readout unit was able to provide the pH value until the new buffer is refilled in the beaker. The rise time and fall time of the sensor for two pH buffers were then evaluated, Figure \ref{fig:5}b represents the changes in sensor response from pH 6.48 to 6.96 and vice versa. The reversibility of the sensor was tested for four cycles between pH 6.48 and 6.96 and plotted in Figure \ref{fig:5}c. It can be noted that the sensor response time in the forward direction is faster than in the reverse direction. The rise time of the sensor is measured to be 11.8 $\pm$ 1.45 s and the fall time was measured to be 21.26 $\pm$ 5.35 s among the four sets of measurements. This asymmetrical behavior is attributed to the rate of diffusion of the analytes into and out of the fluorescence sensor membrane\cite{AN9921701461}.     

\subsection*{Integration of fibre sensors with microfluidic device}
\begin{figure}[ht]
\centering
\includegraphics[width=\linewidth]{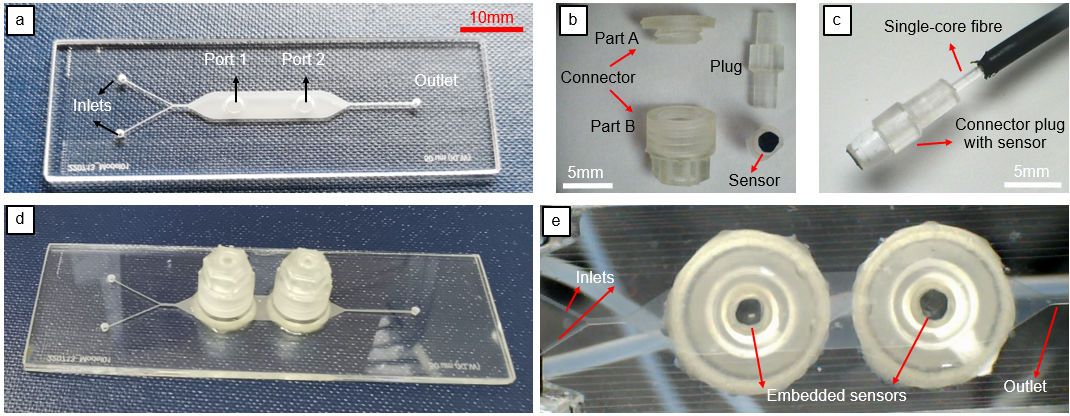}
\caption{Laser-fabricated micromodel and 3D-printed connectors used in the experiment. (a) Micromodel with round ports. \newline (b) 3D-printed connector, connector plug, and connector plug with the attached sensor. (c) Connector plug attached to the commercial interrogation system fibre. (d) Top view of the micromodel with all components attached. (e) Bottom view of the micromodel.}
\label{fig:6}
\end{figure}

To compare the commercial single-core and custom-developed MCF pH systems' performance at the pore scale, both systems were integrated with a laser-fabricated microfluidic device. The device is manufactured from two borosilicate glass plates (Borofloat33, SCHOTT Technical Glass Solutions GmbH, Germany) with dimensions of  75  mm  ×  25  mm  ×  1.1  mm as shown in Figure \ref{fig:6}a, following a fabrication procedure explained previously in literature \cite{mi9080409,Wlodarczyk}. Briefly, a 35 W ultrashort pulse laser (Tangerine, Amplitude-Systemes, France) was used to drill the inlets, outlet, two round ports of diameter 2.8 mm in the first glass plate, and the microfluidic channel (25 mm long and 5 mm wide, 100 $\mu$m deep) in the second glass plate by ablating the material. The two inlets are connected to a Y-shaped channel on the same plane. The same laser was used to weld the first glass plate (with holes) and the second glass plate (with the channels) together, which permanently close the channels.

In order to incorporate the sensors, single-core, and multicore fibres into the microfluidic device, custom connectors were designed and manufactured from clear photocurable resin (FLGPCL04, Formlabs Inc., Somerville, MA, USA) using a desktop, stereolithography 3D printer (Form 2, Formlabs, USA). A similar approach using the 3D printed connectors is previously reported \cite{s21227493}. As illustrated in Figure \ref{fig:6}b, the custom connector consists of two parts, part A and part B, and a  connector plug. The inner diameters of the connector plugs are set to be 1 mm and 1.5 mm to insert the fibres for the single-core (1 mm diameter) and MCF (1.5 mm diameter) systems, respectively. The sensor spot is cut manually to match the outer diameter of the connector plugs and attached to the plug using silicon glue (left overnight to harden) as shown in Figure \ref{fig:6}b. Part A of the connector with the internal thread is attached to the two ports of the micromodel using adhesive (Araldite Standard, Huntsman Advanced Materials  BVBA,  Belgium), and the connector plug with the attached sensor is secured in the ports using the part B of the connector. To ensure a tight connection and avoid any fluid leakage through the ports, the connector plug is wrapped using PTFE tape before inserting the sensor plugs into the micromodels as shown in Figure \ref{fig:6}c. Figure \ref{fig:6}(d)\&(e) shows the top and bottom view of the micromodel with all the components attached.    

\subsubsection*{pH measurements of commercial single-core system in micromodel}
 
Both sensor plugs were inserted into the micromodel for testing. The single-core fibre (PyroScience, 1 mm diameter) (with the protective coating removed, see Figure \ref{fig:6}c) is inserted into port 1 of the device, and pH measurements of various buffer solutions were performed. The MCF fibre (1.5 mm diameter) is inserted in Port 2 and unused during the testing of the single-core system. The inlets were connected to plastic Luer-lock syringes using a PTFE tube. The experimental system allowed us to independently inject buffers in two inlets using two syringe~pumps (AL-300, World Precision Instruments, Sarasota, FL, USA) 
\begin{figure}[http]
\centering
\includegraphics[width=\linewidth]{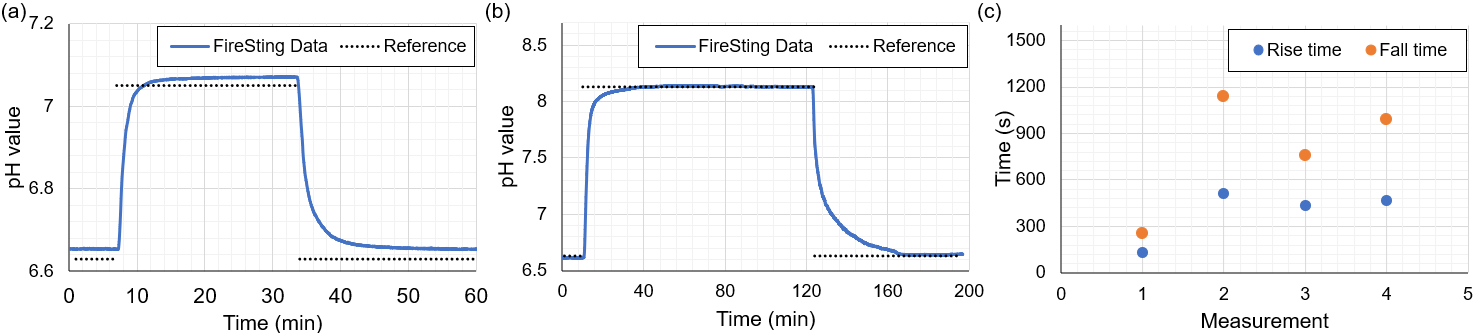}
\caption{pH response behavior of the sensor measured at a flow rate of 150 $\mu$L/min. (a) From pH 6.63 to 7.05 and back to pH 6.63. (b) From pH 6.63 to 8.13 and back to pH 6.63. Dotted lines in (a) and (b) represent the measured pH value with the reference pH meter. (c) Sensor rise times and fall times for four cycles of pH 6.63 and pH 7.05 at a flow rate of 150 $\mu$L/min.}
\label{fig:7}
\end{figure}
\begin{figure}[http]
\centering
\includegraphics[width= \linewidth]{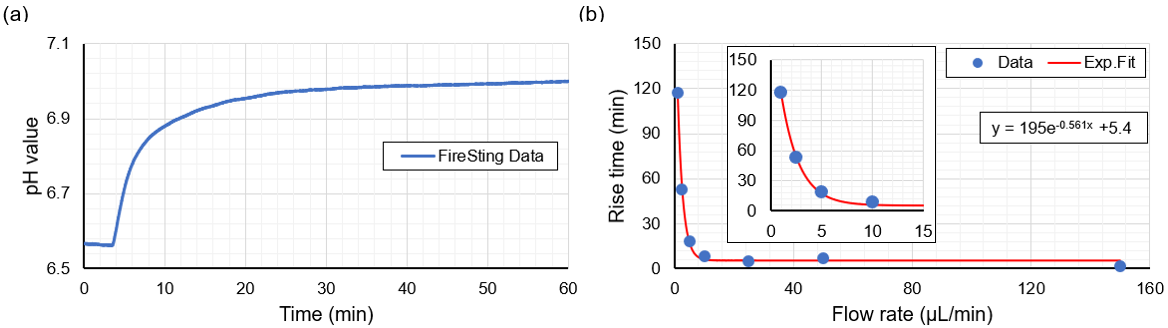}
\caption{(a) Recorded pH values for pH 6.63 to 7.05 for a flow rate of 1 $\mu$L/min. (b) Signal rise time as the function of flow rate and the fit to the data for pH 6.63 and 7.05. Inset- graph shows the first four data points with the fit for better visibility.}
\label{fig:9}
\end{figure}

\hspace{-5 mm}with the dispensing accuracy of $\pm$ 1\%. This ensured the efficient switching of buffers between the pH measurements. The flow of buffers through the microfluidic channel was monitored using a 5-megapixel USB camera. Three buffer solutions were prepared, with pH values of 6.63, 7.05, and 8.13. For the first set of measurements, the syringes were filled with pH 6.63 and 7.05. Initially, pH 6.63 is injected into the microfluidic channel with a rate of 150 $\mu$L/min. Upon fluid flow, the FireSting readout unit started to provide the measured pH values from the sensor spot. The measurement continued until stable pH values (no observable changes in pH measured above the background noise) were obtained in the readout. After obtaining a flat line in the pH value, the injection of the pH 6.63 buffer was stopped, and the second buffer (pH 7.05) was injected into the microfluidic channel through the second inlet with the same flow rate. The sensor responded to the change in the fluid environment and the change in the pH is recorded. The pH value started to increase as the new pH solution replaces the existing pH buffer from the channel. The fluid injection continued until the intensity values were stabilised in the readout, indicating the complete replacement of the pH 6.63 buffer from the channel. After a flat pH readout is obtained, the injection of the fluid is stopped (pH 7.05 buffer), and the solution of pH 6.63 is re-injected into the channel until the pH value is stabilised to the expected value. This allowed us to estimate the signal rise time and fall time of the sensor for a flow rate of 150 $\mu$L/min. The signal rise time (defined as a 90 \% increase from pH 6.63 to pH 7.05) was measured to be 135 s, and the fall time (defined as a 90 \% decrease from pH 7.05 to 6.63) was measured to be 257 s. The measurement is repeated for buffer solutions of pH 6.63 and 8.13 at a flow rate of 150 $\mu$L/min. The signal rise time and fall time of the measurement are determined to be 291 s and 1262 s respectively.   Figure \ref{fig:7}(a)\&(b) represents the pH response curves recorded by the sensor for the two measurements. The reversibility of the sensor in the micromodel was tested for four cycles between pH 6.44 and 7.03 and plotted in Figure \ref{fig:7}c. The rise time was measured to be 387$\pm$ 148 s among four sets, and the fall time was measured to be 786$\pm$ 334 s. Compared with the sensor response time of the order of a few tens of seconds, this highlights the complexity of the micromodel flow. As expected, the fall time is longer than the rise time in all the measurements, similar to the previous calibration tests.

To showcase the feasibility of the single core pH system in a slow pH changing regime, measurements were conducted at a slow flow rate (1 $\mu$L/min). Figure \ref{fig:9}a shows the typical pH readout positive curve for pH 6.63 and pH 7.05. The signal rise time at this flow rate is measured to be 200 minutes. The measurement is repeated for various flow rates and the signal rise time in each measurement is estimated. Figure \ref{fig:9}b shows a plot of signal rise time as a function of flow rate and the exponential fit to the data, which can be used to predict the sensor behavior at various flow rates.  
\subsubsection*{pH measurements of MCF-sensor system in micromodel}

The measurements with the commercial single-core fibre interrogation system provide crucial information about the changes in pH in a micro-environment. Such a system is ideal for the real-time monitoring of the in-situ chemical reactions and process conditions (e.g., cell growth, fluid mix, microreactions). These
\begin{figure}[t!]
\centering
\includegraphics[width= 0.8\linewidth]{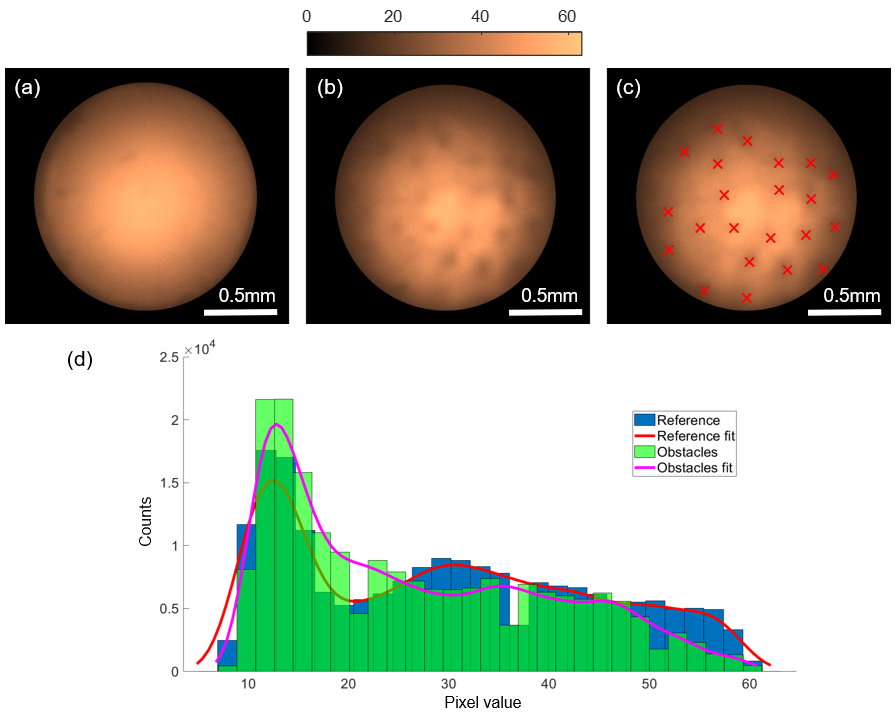}
\caption{False color images of the MCF sensor signal for pH 8.13. (a) Reference image before the obstacles filled in port 2 (b) Image with the microbeads filled in port 2. (c) Image with locations of the obstacles marked with red cross. (d) Histogram plots of reference MCF image and the MCF image in the presence of obstacles with normal distribution fit.}
\label{fig:10}
\end{figure}
measurements can also be conducted with the MCF fibre system, despite the low pH resolution. This is particularly important in a slow-reaction environment as the fibre system will provide the spatial information of the reaction sites within the micromodel. With the current system, we were unable to demonstrate a real-time pH measurement at a slower flow rate due to the potentially large data volume generated by the imaging system. In the future, with on-the-fly image processing software, the measurements can be performed in real-time with the additional benefit of spatial information about the reaction centers within the micromodel, which cannot be offered by any existing commercial single-core fibre pH measurement system.

\begin{figure}[t!]
\centering
\includegraphics[width= \linewidth]{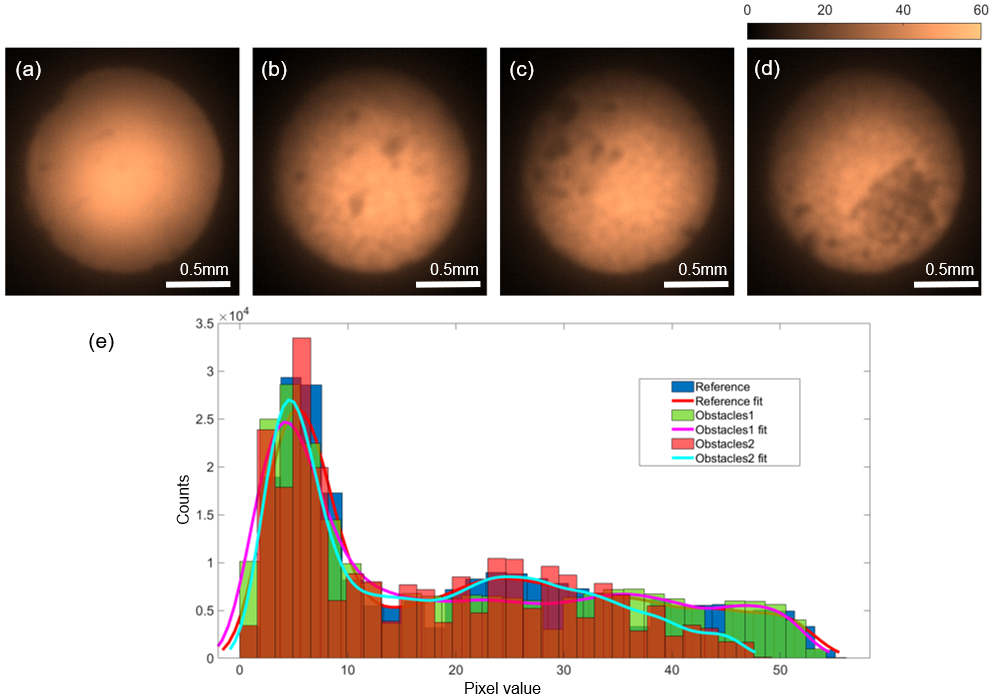}
\caption{False color images of the MCF sensor signal for pH 8.13 (a) Reference image before the rockbeads filled in port 2. (b-c) With the rockbeads inserted for three separate measurements. Note that, the feature shape can be can clearly visible in the MCF images. The smallest feature possesses a dimension of  $\approx$ 50 $\mu$m$\times$ 50 $\mu$m (in Fig b) and the largest feature possesses a dimension of $\approx$ 670 $\mu$m $\times$ 500 $\mu$m (in d).  (e) Histogram plots of reference MCF image (Fig a) and the MCF image in the presence of obstacles (Fig b, labeled as obstacles1) and (Fig d, labeled as obstacles2) with normal distribution fit.}
\label{fig:11}
\end{figure}

To showcase the advantage of the custom-made MCF system over existing single-core pH sensing systems on location-specific pH mapping, proof-of-concept experiments were conducted with the imaging fibre system in the presence of obstacles. First, the imaging fibre is inserted in the mini-Luer connector in port 2 (sensor spot attached). A buffer solution of pH 8.13 is injected into the microchannel using the syringe pump at a flow rate of 150 $\mu$L/min. From the commercial interrogation system measurements, we assessed that the sensor signal rise time is around 300 s in the micromodel. The intensity images from the MCF sensor were recorded after 1500 s of fluid flow, and it is reasonable to assume that the sensor emission stabilised after this time. The sensor plug (and the fibre) is then disconnected from port 2, and microbeads (200 $\mu$m diameter) are embedded in the port, forming a layer of microstructures. This effectively emulates a porous medium consisting of obstacles in the active measurement region. The mini-Luer connector and the fibre is then re-inserted into the system, this time, the sensor was in close contact with the microbeads. An image is taken after injecting the solution of pH 8.13 after 1500 s. Remarkably, the imaging fibre clearly distinguishes the presence of obstacles compared to the images taken without the obstacles present, which can be clearly observed in the emission profile of the fibre sensor. The obstacle restricts the flow of the fluid to the sensor in regions of close contact, hence the fluorescence emission, which in turn reduces the total fluorescence signal from the sensor. The experiment is repeated 4 times by disconnecting and reconnecting the sensor plug and the fibre, showcasing the reversibility of the MCF fibre sensor (see Supplementary Figure S2). It can be noted that, during each set of images, the obstacle's position changed due to slight changes in sensor orientation.

Figure \ref{fig:10}a shows the reference image before the obstacles filled in port 2 and Figure \ref{fig:10}b shows an image with the microbeads filled in port 2. A reduction of 11\% in emission intensity is observed in the image compared to the reference image. The marked points in Figure \ref{fig:10}c indicate the locations of the obstacles which are in contact with the fibre sensor. The reduction in the intensity due to the presence of obstacles can be clearly seen from the line profile across the sensor spot (see Supplementary Figure S3). The change in emission profiles in two measurements can be observed in the histogram of pixel values (Figure \ref{fig:10}d). In the case of measurement with obstacles, the pixels with higher counts get reduced compared to the reference image without the presence of obstacles. Note that, commercial single-core fibre systems can only provide an averaged intensity value for pH evaluation, therefore cannot provide any spatial information about the pH environment. A measurement error (11\% in this case) will occur if any obstacles/particles are in close contact with the sensor spot during the measurements with the commercial single-core fibre system. The MCF system will therefore open a new route toward future pH sensor systems that require spatial pH information in porous structures.    
 
To simulate the conditions in natural porous structures with irregular particle features, measurements were repeated with rock beads. For this, the microbeads, mini-Luer connector, and the MCF were removed from port 2, and rock beads were deposited in port 2. The sensor and the MCF are reinserted into the port and the measurements were repeated several times. Three intensity images are shown in Figure \ref{fig:11}(b-d), and the regions where the rock particles make contact with the sensor are clearly visible among the images, and a reduction of fluorescence emission is observed from the total intensity across the MCF facet. Note that the pixel values vary significantly in the histogram plots of the images (Figure \ref{fig:11}e), where the pixel values are reduced significantly for the case of larger obstacles (Figure \ref{fig:11}d) compared to that of obstacles with smaller features (Figure \ref{fig:11}b). The total counts in Figure \ref{fig:11}b reduced by 4\% compared to the reference image (Figure \ref{fig:11}a) and the total counts in Figure \ref{fig:11}d reduced by 16\% compared to the reference image.

\section*{Conclusion}

We have developed a pH mapping fibre probe using a polymer multicore fibre, that can be integrated with custom-made microfluidic devices and commercially available sensors. The performance of the developed pH fibre sensor in bulk and in micromodels is evaluated and compared with a commercial fibre interrogation system. The commercial single-core and the developed MCF fibre systems are integrated with a laser-fabricated microfluidic device using custom bespoke 3D connectors, manufactured using a stereolithography 3D printer. The connectors allow the integration of sensors within the microfluidic device in a reversible manner, providing experimental flexibility. Exploiting the individual cores of the MCF as independent measurement channels, we demonstrate that the developed fibre sensor can spatially map the pH with tens of micrometer resolution. In order to showcase the feasibility of the developed fibre sensor, we emulate a porous medium by embedding the sensor measurement region within the microfluidic channel with microparticles. The developed fibre system clearly distinguished the particle features that are in contact with the sensor during the fluid flow. Compared to existing commercial pH meters that rely on single-core fibres, the developed fibre system will open a new route toward future pH meters that can provide location-specific parameter measurements for various applications. 

\section*{Data availability}
The data that support the findings of this study are openly available in the Heriot-Watt University PURE research data management system\cite{HKC2023}.

\section*{Acknowledgements}

This research was funded by the European Research Council (ERC) under the European Union’s Horizon 2020 research and innovation programme (MILEPOST, Grant agreement no.: 695070). This paper reflects only the authors’ view and ERC is not responsible for any use that may be made of the information it contains.

\section*{Author contributions statement}

H.K.C proposed the MCF-sensor concept and developed the method with K.L.M, W.N.M, and M.M.M.-V. H.K.C. characterised the experimental system, performed the measurements, analysed the data, and validated outcomes with K.L.W, W.N.M, and M.M.M.-V. K.L.W designed and fabricated the micromodels and the 3D connectors. Resources and supervision by W.N.M and M.M.M.-V. The manuscript was written by H.K.C and further reviewed and edited by W.N.M, K.L.W, and M.M.M.-V. Funding acquisition and project administration by M.M.M.-V.

\section*{Additional information}

{\textbf{Competing interests}}: The authors declare that they have no competing interests. 
\end{document}